\documentclass[a4paper,12pt]{article}
\usepackage{amsmath, amssymb, latexsym, amscd, amsthm,amsfonts,amstext}
\usepackage[mathscr]{eucal}
\usepackage{graphicx}
\usepackage{subfig}

\numberwithin{equation}{section}
\input{tcilatex}
\begin{document}

\title{ Globally strictly convex cost functional for an inverse parabolic
problem}
\author{Michael V. Klibanov$^{\ast }$ and Vladimir G. Kamburg$^{\circ }$
\and $^{\ast }$Department of Mathematics and Statistics \and University of
North Carolina at Charlotte \and Charlotte, NC 28223, USA \and $^{\circ }$%
Department of Information and Computing Systems \and Penza State University
of Acrhitecture and Construction, \and Penza 440028, Russian Federation \and %
E-mails: mklibanv@uncc.edu and kamburg@rambler.ru}
\date{}
\maketitle

\begin{abstract}
A coefficient inverse problem for a parabolic equation is considered. Using
a Carleman Weight Function, a globally strictly convex cost functional is
constructed for this problem.
\end{abstract}



\graphicspath{{Figures/}}

%
%
%
%
%
%
%
%
%
%
%

%
%

\graphicspath{{FIGURES/}
{Figures/}
{FiguresJ/newfigures/}
{pics/}}

\section{Introduction}

\label{sec:1}

Both the most challenging and the most important question one needs to
address when trying to solve numerically a Coefficient Inverse Problem (CIP)
is: \emph{How to obtain a rigorously guaranteed good approximation for the
exact solution without any advanced knowledge of a small neighborhood of
this solution?} We call a numerical method addressing this question \emph{%
globally convergent}. The reason of the importance of this question is that
conventional least squares cost functionals for CIPs are non-convex. Hence,
they have many local minima and ravines. This makes numerical procedures,
which use these functionals, unreliable.\ Indeed, to find a proper minimum,
one should start the minimization process in a small neighborhood of the
exact solution, i.e. one should start from a good first guess for the
solution. However, if such a guess is known a priori, then why the exact
solution is also not known a priori? In addition, such \ a neighborhood is
very rarely available in practice.\ Therefore, it is important to develop
globally convergent numerical methods for CIPs.

Currently there exist two types of such methods. Both for the case of the
non overdetermined data, i.e. for the case of a single measurement event.
The first type is based on constructions of globally strictly convex
Tikhonov-like functionals. Carleman Weight Functions (CWFs) are the key to
the global convexity. This method was initiated in works of Klibanov in 1997 
\cite{Klib97,Nonl}; also see Klibanov and Timonov \cite{KT}. Recently there
is a renewed interest in Beilina and Klibanov \cite{BKnonl} and Klibanov and
Th\'{a}nh \cite{KTnonl} with some numerical studies in \cite{KTnonl}. A
different, although a similar approach, was carried out by Baudouin, De
Buhan and Ervedoza \cite{Bad}. The global convexity is understood as
follows: Given a convex set $G$ of an arbitrary diameter $d$ in a certain
Sobolev space, one can choose the parameter $\lambda _{0}=\lambda _{0}\left(
d\right) >>1$ of the CWF such that for all $\lambda \geq \lambda _{0}$ that
functional is strictly convex on $K$. Assume now that there exists a
minimizer of that functional on the set $K$. Then the strict convexity
guarantees convergence of the gradient method to this minimizer starting
from any point of the set $K$ \cite{BKnonl,KTnonl}. This is the global
convergence as in the above definition. Recall that in the conventional case
the gradient method converges to a minimizer only if starting in a small
neighborhood of that minimizer.

The second type of globally convergent numerical methods for CIPs is the
method, which was initiated in the paper of Beilina and Klibanov \cite{BK1}
and was discussed since then in a number of follow up publications of these
authors with coauthors. Results obtained before 2012 were summarized in the
book \cite{BK2}. In particular, this method was completely verified on
experimental data, see, e.g., Chapter 5 in \cite{BK2}, as well as \cite%
{BK3,BTKF,TBKF}.\ We also refer to a recent paper of Chow and Zou \cite{Chow}
for this method.

The method of this paper falls into category of the first type of globally
convergent numerical methods. Using a CWF for the parabolic operator, we
construct a globally strictly convex cost functional for a CIP for a general
parabolic equation of the second order. Unlike this, Note that CIPs for
hyperbolic PDEs were considered in \cite{Bad,BKnonl,Klib97,KTnonl}. Although
a CIP for a parabolic PDE was considered in \cite{Nonl}, the main difference
of that work with the current paper is that in \cite{Nonl} a certain series
was truncated, whereas truncation does not take place here.

In section 2 we state our inverse problem. The globally strictly convex cost
functional for it is constructed in section 3. In section 4 we prove the
main result of this paper, which is Theorem 1.

\section{Statement of the problem}

\label{sec:2}

Let $\Omega \subset \mathbb{R}^{3}$ be a bounded domain with a piecewise
smooth boundary $\partial \Omega .$ Let $T>0$ be an arbitrary number.\
Denote $\Omega _{T}=\Omega \times \left( -T,T\right) .$ Let $\Gamma
\subseteq \partial \Omega $ be a part of the boundary $\partial \Omega $ and
let $\Gamma \in C^{2}.$ Denote $\Gamma _{T}=\Gamma \times \left( -T,T\right)
.$ Consider the elliptic operator of the second order in $\Omega ,$%
\begin{equation}
Lu=\dsum\limits_{i,j=1}^{3}a_{ij}\left( x\right)
u_{x_{i}x_{j}}+\dsum\limits_{i,j=1}^{3}b_{j}\left( x\right)
u_{x_{j}}+c\left( x\right) u,x\in \Omega .  \label{2.1}
\end{equation}%
Here 
\begin{equation}
a_{ij}=a_{ji},a_{ij}\in C^{1}\left( \overline{\Omega }\right) ;b_{j},c\in
C\left( \overline{\Omega }\right) ,  \label{2.2}
\end{equation}%
\begin{equation}
\mu _{1}\left\vert \xi \right\vert ^{2}\leq
\dsum\limits_{i,j=1}^{3}a_{ij}\left( x\right) \xi _{i}\xi _{j}\leq \mu
_{2}\left\vert \xi \right\vert ^{2},\forall x\in \overline{\Omega },\forall
\xi \in \mathbb{R}^{n},  \label{2.3}
\end{equation}%
where $\mu _{1},\mu _{2}=const.>0,\mu _{1}\leq \mu _{2}.$ Let the function $%
u\in C^{4,2}\left( \overline{\Omega }_{T}\right) $ satisfies the following
conditions%
\begin{equation}
u_{t}=Lu\text{ in }\Omega _{T},  \label{2.4}
\end{equation}%
\begin{equation}
u\left( x,0\right) =f\left( x\right) .  \label{2.5}
\end{equation}

We assume here that equation (\ref{2.4}) is valid not only for $t>0$, but
for $t<0$ as well. This is because the Bukhgeim-Klibanov method \cite%
{BukhKlib} does not work for the case when a parabolic equation is valid
only for $t>0$ \cite{BK2,KT,K}. Our interest is in the inverse problem which
we now formulate.

\textbf{Coefficient Inverse Problem (CIP).} \emph{Assume that the
coefficient }$c\left( x\right) $\emph{\ in (\ref{2.1}) is unknown for }$x\in
\Omega .$\emph{\ On the other hand, assume that the following functions }$%
g_{1}\left( x,t\right) ,g_{2}\left( x,t\right) $\emph{\ are known,}%
\begin{equation}
u\mid _{\Gamma _{T}}=g_{1}\left( x,t\right) ,\partial _{n}u\mid _{\Gamma
_{T}}=g_{2}\left( x,t\right) .  \label{2.6}
\end{equation}%
\emph{Determine }$c\left( x\right) $\emph{\ for }$x\in \Omega .$

Assume that in (\ref{2.5})%
\begin{equation}
f\left( x\right) \geq 2b=const.>0\text{ in }\overline{\Omega }^{d}.
\label{2.7}
\end{equation}%
Then uniqueness of this CIP follows immediately from Theorem 1.10.7 of \cite%
{BK2} as well as from Theorem 3.3.2 of \cite{KT} and Theorem 3.4 of \cite{K}%
. Note that a particular case of the data (\ref{2.6}) is the case of
backscattering data. The principal parts of the operators $L$ and $\partial
_{t}-L$ we denote as $L_{0},P_{0},$%
\begin{equation}
L_{0}u=\dsum\limits_{i,j=1}^{3}a_{ij}\left( x\right)
u_{x_{i}x_{j}},P_{0}u=u_{t}-L_{0}u.  \label{2.8}
\end{equation}

\section{The cost functional}

\label{sec:3}

To construct a globally convergent numerical method for our CIP, we
construct in this section a globally strictly convex cost functional. Below $%
x=\left( x_{1},\overline{x}\right) ,$ where $\overline{x}=\left(
x_{2},x_{3}\right) .$ Without any loss of generality we can assume that $%
\Gamma =\left\{ x:x_{1}=p\left( \overline{x}\right) ,p\in C^{2}\left(
\left\vert \overline{x}\right\vert \leq \sqrt{d}\right) \right\} $ for a
number $d\in \left( 0,1\right) .$ Therefore, changing variables as $\left(
x_{1},\overline{x}\right) \Leftrightarrow \left( x_{1}^{\prime },\overline{x}%
\right) ,$ where $x_{1}^{\prime }:=x_{1}-p\left( \overline{x}\right) ,$ and
using the same notations as before, for brevity, we conclude that we can
assume that%
\begin{equation}
\Gamma =\left\{ x:x=\left( 0,\overline{x}\right) ,\left\vert \overline{x}%
\right\vert <\sqrt{d}\right\} .  \label{3.1}
\end{equation}%
Denote%
\begin{equation}
\Omega ^{d}=\left\{ x:x_{1}>0,\left\vert \overline{x}\right\vert <\sqrt{d}%
\right\}  \label{3.2}
\end{equation}%
and assume below that $\Omega ^{d}\subset \Omega .$ Then $\Gamma \subset
\partial \Omega ^{d}.$ Below we determine the unknown coefficient $c\left(
x\right) $ only in a subdomain of the domain $\Omega ^{d}.$

We now formulate the Carleman estimate for the operator $P_{0}=\partial
_{t}-L_{0}.$ Let $\lambda >1$ and $\nu >1$ be two large parameters, which we
define later. Consider an arbitrary number $a\in \left( 0,d\right) .$
Consider functions $\psi \left( x,t\right) $, $\varphi _{\lambda }\left(
x,t\right) $ defined as%
\begin{equation*}
\psi \left( x,t\right) =x_{1}+\left\vert \overline{x}\right\vert ^{2}+\frac{%
t^{2}}{T^{2}}+a,\text{ }\varphi _{\lambda }\left( x,t\right) =\exp \left(
\lambda \psi ^{-\nu }\right) .
\end{equation*}%
Consider the following sets%
\begin{equation*}
G_{a,d}=\left\{ \left( x,t\right) :x_{1}>0,x_{1}+\left\vert \overline{x}%
\right\vert ^{2}+\frac{t^{2}}{T^{2}}+a<d\right\} ,
\end{equation*}%
\begin{equation*}
G_{a,d}^{0}=G_{a,d}\cap \left\{ t=0\right\} ,
\end{equation*}%
\begin{equation*}
\xi _{a,d}=\left\{ \left( x,t\right) :x_{1}>0,x_{1}+\left\vert \overline{x}%
\right\vert ^{2}+\frac{t^{2}}{T^{2}}+a=d\right\} ,
\end{equation*}%
\begin{equation*}
\Gamma _{a,d,T}=\left\{ \left( x,t\right) :x_{1}=0,\left\vert \overline{x}%
\right\vert ^{2}+\frac{t^{2}}{T^{2}}<d-a\right\} \subset \Gamma _{T},
\end{equation*}%
\begin{equation}
\partial G_{a.d}=\xi _{a,d}\cup \Gamma _{a,d,T},  \label{3.6}
\end{equation}%
\begin{equation*}
G_{a,d-\varepsilon }=\left\{ \left( x,t\right) :x_{1}>0,x_{1}+\left\vert 
\overline{x}\right\vert ^{2}+\frac{t^{2}}{T^{2}}+a<d-\varepsilon \right\} ,
\end{equation*}%
where $\varepsilon \in \left( 0,d-a\right) $ is a sufficiently small number.
Clearly $G_{a,d-\varepsilon }\subset G_{a,d}.$ Since $d\in \left( 0,1\right)
,$ then $G_{a,d}\subset \Omega _{T},G_{a,d}^{0}\subset \Omega ^{d}$ Also, $%
\xi _{a}$ is the level surface of both functions $\psi ,\varphi _{\lambda }$%
. Note that $\xi _{a,d}$ is the level surface of the function $\varphi
_{\lambda },$%
\begin{equation}
\min_{\overline{G}_{a,d}}\varphi _{\lambda }^{2}=\varphi _{\lambda }^{2}\mid
_{\xi _{a,d}}=\exp \left[ 2\lambda d^{-\nu }\right] .  \label{3.8}
\end{equation}%
Lemma 1 follows immediately from Lemma 3 of \S 1 of chapter 4 of the book of
Lavrentiev, Romanov and Shishatskii \cite{LRS}.

\textbf{Lemma 1 }(Carleman estimate). \emph{There exist sufficiently large
numbers }$\nu _{0},\lambda _{0}>1,$\emph{\ }%
\begin{equation*}
\nu _{0}=\nu _{0}\left( a,d,\mu _{1},\mu _{2},\max_{i,j}\left\Vert
a_{i,j}\right\Vert _{C^{1}\left( \overline{\Omega }^{d}\right) },T\right)
,\lambda _{0}=\lambda _{0}\left( a,d,\mu _{1},\mu _{2},\max_{i,j}\left\Vert
a_{i,j}\right\Vert _{C^{1}\left( \overline{\Omega }^{d}\right) },T\right)
\end{equation*}%
\emph{\ \ depending only on listed parameters and a sufficiently large
absolute constant }$\lambda _{0}>1$\emph{\ such that for all }$\nu \geq \nu
_{0},\lambda \geq \lambda _{0}$\emph{\ and for all functions }$u\in
C^{2,1}\left( \overline{G}_{a,d}\right) $\emph{\ the following pointwise
Carleman estimate is valid for all }$\left( x,t\right) \in G_{a,d}$\emph{\ }%
\begin{eqnarray*}
\left( P_{0}u\right) ^{2}\varphi _{\lambda }^{2} &\geq &C\lambda \left\vert
\nabla u\right\vert ^{2}\varphi _{\lambda }^{2}+C\lambda ^{3}u^{2}\varphi
_{\lambda }^{2}+\func{div}U+V_{t}, \\
\left\vert U\right\vert ,\left\vert V\right\vert &\leq &C\lambda ^{3}\left[
\left( \nabla u\right) ^{2}+u_{t}^{2}+u^{2}\right] \varphi _{\lambda }^{2},
\end{eqnarray*}%
\emph{where the constant }$C=C\left( n,\max_{i,j}\left\Vert
a_{i,j}\right\Vert _{C^{1}\left( \overline{\Omega }^{d}\right) }\right) >0$%
\emph{\ depends only on listed parameters.}

Denote%
\begin{equation*}
L_{c}u=Lu-c\left( x\right) u=\dsum\limits_{i,j=1}^{3}a_{ij}\left( x\right)
u_{x_{i}x_{j}}+\dsum\limits_{i,j=1}^{3}b_{j}\left( x\right) u_{x_{j}}.
\end{equation*}%
Since the function $u\in C^{4,2}\left( \overline{\Omega }_{T}\right) ,$ then
(\ref{2.7}) implies that $u\left( x,t\right) \geq b>0$ in $\overline{G}%
_{a,d} $ for sufficiently small $T.$ Hence, we can consider the function $%
v\left( x,t\right) =\ln u\left( x,t\right) .$ Substituting $u=e^{v}$ in (\ref%
{2.4}) and (\ref{2.5}), we obtain%
\begin{equation}
v_{t}=L_{c}v+\dsum\limits_{i,j=1}^{n}a_{i,j}\left( x\right)
v_{x_{i}}v_{x_{j}}+c\left( x\right) \text{ in }\overline{G}_{a,d},
\label{3.10}
\end{equation}%
\begin{equation}
v\left( x,0\right) =\ln f\left( x\right) .  \label{3.11}
\end{equation}%
Let $w\left( x,t\right) =v_{t}\left( x,t\right) .$ Differentiate (\ref{3.10}%
) with respect to $t$ and use (\ref{3.11}). We obtain the following
nonlinear integral differential equation with respect to the function $w$ in
the domain in $\overline{G}_{a,d}$%
\begin{equation*}
w_{t}=L_{c}w+
\end{equation*}%
\begin{equation}
+\dsum\limits_{i,j=1}^{n}a_{i,j}\left( x\right) w_{x_{i}}\left( \left( \ln
f\right) _{x_{j}}+\dint\limits_{0}^{t}w_{x_{j}}\left( x,\tau \right) d\tau
\right)  \label{3.12}
\end{equation}%
\begin{equation*}
+\dsum\limits_{i,j=1}^{n}a_{i,j}\left( x\right) w_{x_{j}}\left( \left( \ln
f\right) _{x_{i}}+\dint\limits_{0}^{t}w_{x_{i}}\left( x,\tau \right) d\tau
\right) .
\end{equation*}%
In addition, conditions (\ref{2.6}) imply that 
\begin{equation}
w\mid _{\Gamma _{a,d,T}}=\widetilde{g}_{1}\left( x,t\right) ,\partial
_{x_{1}}w\mid _{\Gamma _{a,d,T}}=\widetilde{g}_{2}\left( x,t\right) ,
\label{3.13}
\end{equation}%
where 
\begin{equation*}
\widetilde{g}_{1}\left( x,t\right) =\partial _{t}\ln g_{1}\left( x,t\right) ,%
\widetilde{g}_{2}\left( x,t\right) =\frac{g_{2t}}{g_{1}}-\frac{g_{1t}g_{2}}{%
g_{1}^{2}}.
\end{equation*}%
Since functions $g_{1},g_{2}$ are the data for the inverse problem, then
they naturally contain noise. Even though the differentiation of a noisy
function is an ill-posed problem, it can be handled by a number of well
known regularization methods, see, e.g. Aristov \cite{A}.

Thus, we have obtained the nonlinear integral differential equation (\ref%
{3.12}) with the lateral Cauchy data (\ref{3.13}). If we find the solution
of this problem, then backwards calculations will deliver us the target
coefficient $c\left( x\right) $ for $x\in G_{a,d}^{0}.$ Hence, we focus
below on the solution of the problem (\ref{3.12}), (\ref{3.13}). Denote $%
\widetilde{L}w$ the right hand side of (\ref{3.12}), plus the term $-w_{t}$,%
\begin{equation*}
\widetilde{L}w=-w_{t}+L_{c}w+
\end{equation*}%
\begin{equation}
+\dsum\limits_{i,j=1}^{n}a_{i,j}\left( x\right) w_{x_{i}}\left( \left( \ln
f\right) _{x_{j}}+\dint\limits_{0}^{t}w_{x_{j}}\left( x,\tau \right) d\tau
\right)  \label{3.14}
\end{equation}%
\begin{equation*}
+\dsum\limits_{i,j=1}^{n}a_{i,j}\left( x\right) w_{x_{j}}\left( \left( \ln
f\right) _{x_{i}}+\dint\limits_{0}^{t}w_{x_{i}}\left( x,\tau \right) d\tau
\right) .
\end{equation*}%
Our weighted Tikhonov-like cost functional is%
\begin{equation}
J_{\lambda ,\alpha }\left( w\right) =\exp \left( -3\lambda d^{-\nu }\right)
\dint\limits_{G_{a,d}}\left( \widetilde{L}w\right) ^{2}\varphi _{\lambda
}^{2}dxdt+\alpha \left\Vert w\right\Vert _{H^{4}\left( G_{a,d}\right) }^{2},
\label{3.15}
\end{equation}%
where $\alpha \in \left( 0,1\right) $ is the regularization parameter. We
use the multiplier $\exp \left( -3\lambda d^{-\nu }\right) $ to ensure that
we can indeed choose $\alpha \in \left( 0,1\right) ,$ see Theorem 1. We use
the $H^{4}\left( G_{a,d}\right) -$ norm here since we need in our proof $%
w\in C^{1}\left( \overline{G}_{a,d}\right) \cap H^{2}\left( G_{a,d}\right) ,$
and the embedding theorem guarantees that 
\begin{equation}
H^{4}\left( G_{a,d}\right) \subset C^{1}\left( \overline{G}_{a,d}\right)
,\left\Vert u\right\Vert _{C^{1}\left( \overline{G}_{a,d}\right) }\leq
C_{1}\left\Vert u\right\Vert _{H^{4}\left( G_{a,d}\right) },\forall u\in
H^{4}\left( G_{a,d}\right) ,  \label{3.16}
\end{equation}%
where $C_{1}=C_{1}\left( G_{a,d}\right) >0$ is a generic constant depending
only on the domain $G_{a,d}.$ Thus, we consider below the following problem.

\textbf{Minimization Problem}. \emph{Minimize the functional }$J_{\lambda
,\alpha }\left( w\right) $\emph{\ in (\ref{3.15}), subject to the lateral
Cauchy data (\ref{3.13}).}

Let $R>0$ be an arbitrary number. Consider the set $B\left( R\right) \subset
H^{4}\left( G_{a,d}\right) ,$%
\begin{equation*}
B\left( R\right) =
\end{equation*}%
\begin{equation}
\left\{ w\in H^{4}\left( G_{a,d}\right) :\left\Vert w\right\Vert
_{H^{4}\left( G_{a,d}\right) }<R,w\text{ satisfies boundary conditions (\ref%
{3.13})}\right\} .  \label{3.17}
\end{equation}%
Introduce the space $H_{0}^{4}\left( G_{a,d}\right) $ as%
\begin{equation*}
H_{0}^{4}\left( G_{a,d}\right) =\left\{ u\in H^{4}\left( G_{a,d}\right)
:u\mid _{\Gamma _{a,d,T}}=\partial _{x_{1}}u\mid _{\Gamma
_{a,d,T}}=0\right\} .
\end{equation*}

\textbf{Theorem 1}. \emph{For all numbers }$\lambda ,\nu ,\alpha >0$\emph{\
and for all functions }$w\in B\left( R\right) $\emph{\ there exists the Frech%
\'{e}t derivative }$J_{\lambda ,\alpha }^{\prime }\left( w\right) \in
H_{0}^{4}\left( G_{a,d}\right) $\emph{\ of the functional }$J_{\lambda
,\alpha }$\emph{\ at the point }$w$\emph{. Let }$\nu =\nu _{0}$\emph{\ be
the sufficiently large number of Lemma 1 and }$b>0$\emph{\ be the number in (%
\ref{2.7}). There exists a sufficiently large number }$\lambda _{1},$\emph{\ 
}%
\begin{equation}
\lambda _{1}=\lambda _{1}\left( a,d,\mu _{1},\mu _{2},\max_{i,j}\left\Vert
a_{i,j}\right\Vert _{C^{1}\left( \overline{\Omega }^{d}\right)
},T,b,R,\left\Vert \nabla f\right\Vert _{C\left( \overline{G}%
_{a,d}^{0}\right) }\right) \geq \lambda _{0}>1  \label{3.18}
\end{equation}%
\emph{depending on listed parameters such that if the regularization
parameter }

$\alpha \in \left( \exp \left( -\lambda /\left( 2d^{\nu }\right) \right)
,1\right) ,$\emph{\ then the functional }$J_{\lambda ,\alpha }\left(
w\right) $\emph{\ is strictly convex on the set }$B\left( R\right) $\emph{\
for all }$\lambda \geq \lambda _{1}.$\emph{\ More precisely, }%
\begin{equation*}
J_{\lambda ,\alpha }\left( w_{2}\right) -J_{\lambda ,\alpha }\left(
w_{1}\right) -J_{\lambda ,\alpha }^{\prime }\left( w_{1}\right) \left(
w_{2}-w_{1}\right) \geq
\end{equation*}%
\begin{equation}
C_{1}\exp \left( 2\lambda q\right) \dint\limits_{G_{a+\varepsilon ,d}}\left[
\left( \nabla w_{2}-\nabla w_{1}\right) ^{2}+\left( w_{2}-w_{1}\right) ^{2}%
\right] dxdt+\frac{\alpha }{2}\left\Vert w_{2}-w_{1}\right\Vert
_{H^{4}\left( G_{a,d}\right) }^{2},  \label{3.19}
\end{equation}%
\begin{equation*}
\forall w_{1},w_{2}\in B\left( R\right) ,\forall \lambda \geq \lambda _{1}.
\end{equation*}%
\emph{where }$q=\left( d-\varepsilon \right) ^{-\nu _{0}}\left[ 1-3\left(
d-\varepsilon \right) ^{\nu _{0}}/\left( 2d\right) ^{\nu _{0}}\right] >0$ 
\emph{and the constant }$C_{1}>0$\emph{\ depends on the same parameters as
those in (\ref{3.18}).}

Note that we can require $\alpha \in \left( \exp \left( -\lambda /\left(
2d^{\nu }\right) \right) ,1\right) ,$ since $\exp \left( -\lambda /\left(
2d^{\nu }\right) \right) <<1$ for sufficiently large $\lambda .$ This
theorem is the main result of our paper. Theorem 1 enables one to prove the
convergence of the gradient method, which can start at any point of the set $%
B\left( R\right) .$ Since there are no restrictions on the diameter $2R$ of
this set, then this is the global convergence as defined in Introduction.
The stability with respect to the noise in the data (\ref{3.13}) can also be
established using this theorem. These two latter results can be derived from
Theorem 1 in the same manner as similar results are derived from global
strict convexity theorems in \cite{BKnonl,KTnonl}. Hence, we do not describe
these results here. Below we focus on the proof of Theorem 1 and assume that
its conditions are satisfied. Below $C_{1}>0$ denotes different constants
depending on the same parameters as ones listed in (\ref{3.18}).

\section{Proof of Theorem 1}

\label{sec:4}

Let $w_{1},w_{2}\in B\left( R\right) $ be two arbitrary functions. Denote $%
h=w_{2}-w_{1}.$\ Then%
\begin{equation}
h\in H_{0}^{4}\left( G_{a,d}\right) ,\left\Vert h\right\Vert _{H^{4}\left(
G_{a,d}\right) }\leq 2R.  \label{4.1}
\end{equation}%
Let $A=\left( \widetilde{L}\left( w_{1}+h\right) \right) ^{2}-\left( 
\widetilde{L}\left( w_{1}\right) \right) ^{2}.$ First, we single out the
linear part of this expression with respect to $h$. Using (\ref{2.1}) and (%
\ref{3.14}), we obtain 
\begin{equation*}
\widetilde{L}\left( w_{1}+h\right) =\widetilde{L}\left( w_{1}\right)
+L_{c}\left( h\right) -h_{t}
\end{equation*}%
\begin{equation*}
+\dsum\limits_{i,j=1}^{3}a_{i,j}\left( x\right) h_{x_{i}}\left( \left( \ln
f\right) _{x_{j}}+\dint\limits_{0}^{t}w_{1x_{j}}\left( x,\tau \right) d\tau
\right) +
\end{equation*}%
\begin{equation*}
+\dsum\limits_{i,j=1}^{3}a_{i,j}\left( x\right) h_{x_{j}}\left( \left( \ln
f\right) _{x_{i}}+\dint\limits_{0}^{t}w_{1x_{i}}\left( x,\tau \right) d\tau
\right) +
\end{equation*}%
\begin{equation*}
+\dsum\limits_{i,j=1}^{3}a_{i,j}\left( x\right) \left[ w_{1x_{i}}\dint%
\limits_{0}^{t}h_{x_{j}}\left( x,\tau \right) d\tau
+w_{1x_{j}}\dint\limits_{0}^{t}h_{x_{i}}\left( x,\tau \right) d\tau \right] .
\end{equation*}%
Hence,%
\begin{equation*}
A=2\widetilde{L}\left( w_{1}\right) \left[ L\left( h\right)
-h_{t}+\dsum\limits_{i,j=1}^{3}a_{i,j}\left( x\right) h_{x_{i}}\left( \left(
\ln f\right) _{x_{j}}+\dint\limits_{0}^{t}w_{1x_{j}}\left( x,\tau \right)
d\tau \right) \right]
\end{equation*}%
\begin{equation*}
+2\widetilde{L}\left( w_{1}\right) \dsum\limits_{i,j=1}^{3}a_{i,j}\left(
x\right) h_{x_{j}}\left( \left( \ln f\right)
_{x_{i}}+\dint\limits_{0}^{t}w_{1x_{i}}\left( x,\tau \right) d\tau \right)
\end{equation*}%
\begin{equation*}
+2\widetilde{L}\left( w_{1}\right) \dsum\limits_{i,j=1}^{3}a_{i,j}\left(
x\right) \left[ w_{1x_{i}}\dint\limits_{0}^{t}h_{x_{j}}\left( x,\tau \right)
d\tau +w_{1x_{j}}\dint\limits_{0}^{t}h_{x_{i}}\left( x,\tau \right) d\tau %
\right] +S^{2}\left( h,w_{1}\right) ,
\end{equation*}%
where 
\begin{equation*}
S\left( h,w_{1}\right) =L\left( h\right) -h_{t}
\end{equation*}%
\begin{equation*}
+\dsum\limits_{i,j=1}^{3}a_{i,j}\left( x\right) h_{x_{i}}\left( \left( \ln
f\right) _{x_{j}}+\dint\limits_{0}^{t}w_{1x_{j}}\left( x,\tau \right) d\tau
\right)
\end{equation*}%
\begin{equation}
+\dsum\limits_{i,j=1}^{3}a_{i,j}\left( x\right) h_{x_{j}}\left( \left( \ln
f\right) _{x_{i}}+\dint\limits_{0}^{t}w_{1x_{i}}\left( x,\tau \right) d\tau
\right)  \label{4.2}
\end{equation}%
\begin{equation*}
+\dsum\limits_{i,j=1}^{3}a_{i,j}\left( x\right) \left[ w_{1x_{i}}\dint%
\limits_{0}^{t}h_{x_{j}}\left( x,\tau \right) d\tau
+w_{1x_{j}}\dint\limits_{0}^{t}h_{x_{i}}\left( x,\tau \right) d\tau \right] .
\end{equation*}%
The expression $D\left( w_{1},h\right) =A-S^{2}\left( h,w_{1}\right) $ is
linear with respect to $h$. Hence, consider the functional 
\begin{equation*}
Q_{\lambda ,\alpha }\left( h\right) =\exp \left( -3\lambda d^{-\nu }\right)
\dint\limits_{G_{a,d}}D\left( w_{1},h\right) \varphi _{\lambda
}^{2}dxdt+2\alpha \left[ h,w_{1}\right] ,
\end{equation*}%
where $\left[ ,\right] $ is the scalar product in $H^{4}\left(
G_{a,d}\right) .$ This is a linear bounded functional acting from $%
H_{0}^{4}\left( G_{a,d}\right) $ into $\mathbb{R}.$ Hence, by Riesz theorem
there exists unique element $U_{\lambda ,\alpha }\left( w_{1}\right) \in
H_{0}^{4}\left( G_{a,d}\right) $ such that $Q_{\lambda ,\alpha }\left(
h\right) =\left[ U_{\lambda ,\alpha }\left( w_{1}\right) ,h\right] $.
Furthermore, the norm $\left\Vert U_{\lambda ,\alpha }\right\Vert
_{H^{4}\left( G_{a,d}\right) }$ equals to the norm of the functional $%
Q_{\lambda ,\alpha }.$ Hence, we have proven the existence of the Frech\'{e}%
t derivative $J_{\lambda ,\alpha }^{\prime }\left( w_{1}\right) =U_{\lambda
,\alpha }\left( w_{1}\right) \in H_{0}^{4}\left( G_{a,d}\right) $ of the
functional $J_{\lambda ,\alpha }$ and 
\begin{equation*}
J_{\lambda ,\alpha }^{\prime }\left( w_{1}\right) \left( h\right) =\exp
\left( -3\lambda d^{-\nu }\right) \dint\limits_{G_{a,d}}B\left(
w_{1},h\right) \varphi _{\lambda }^{2}dxdt+2\alpha \left[ h,w_{1}\right]
,\forall h\in H_{0}^{4}\left( G_{a,d}\right) .
\end{equation*}%
Hence, 
\begin{equation}
J_{\lambda ,\alpha }\left( w_{1}+h\right) -J_{\lambda ,\alpha }\left(
w_{1}\right) -J_{\lambda ,\alpha }^{\prime }\left( w_{1}\right) \left(
h\right)  \label{4.3}
\end{equation}%
\begin{equation*}
=\exp \left( -3\lambda d^{-\nu }\right) \dint\limits_{G_{a,d}}S^{2}\left(
h,w_{1}\right) \varphi _{\lambda }^{2}dxdt+\alpha \left\Vert h\right\Vert
_{H^{4}\left( G_{a,d}\right) }^{2},
\end{equation*}%
where $S\left( h,w_{1}\right) $ is given in (\ref{4.2}).

We now focus on the estimate from the below of the integral in (\ref{4.3}).
Because of Lemma 1, we single out the term with $\left( h_{t}-L\left(
h\right) \right) ^{2}.$ Using (\ref{2.2}), (\ref{2.7}), (\ref{3.14}) and the
Cauchy-Schwarz inequality, we obtain%
\begin{equation*}
S^{2}\left( h,w_{1}\right) \geq \frac{1}{2}\left( h_{t}-L\left( h\right)
\right) ^{2}-C_{1}\left( \nabla h\right) ^{2}-C_{1}\left(
\dint\limits_{0}^{t}\left\vert \nabla h\left( x,\tau \right) \right\vert
^{2}d\tau \right)
\end{equation*}%
\begin{equation}
\geq \frac{1}{3}\left( P_{0}\left( h\right) \right) ^{2}-C_{1}\left( \nabla
h\right) ^{2}-C_{1}\left( \dint\limits_{0}^{t}\left\vert \nabla h\left(
x,\tau \right) \right\vert d\tau \right) ^{2},  \label{4.4}
\end{equation}%
where the operator $P_{0}=\partial _{t}-L_{0}$ was defined in (\ref{2.8}).
It follows from Lemma 1.10.3 of \cite{BK2} that%
\begin{equation}
\dint\limits_{G_{a,d}}\left( \dint\limits_{0}^{t}\left\vert \nabla h\left(
x,\tau \right) \right\vert d\tau \right) ^{2}\varphi _{\lambda }^{2}dxdt\leq 
\frac{C_{1}}{\lambda }\dint\limits_{G_{a,d}}\left( \nabla h\right)
^{2}\varphi _{\lambda }^{2}dxdt.  \label{4.5}
\end{equation}%
Hence, using Lemma 1, (\ref{3.6}), (\ref{3.8}), (\ref{4.1}), (\ref{4.4}) and
(\ref{4.5}), we obtain for sufficiently large $\lambda \geq \lambda _{1}$%
\begin{equation*}
\exp \left( -3\lambda d^{-\nu }\right) \dint\limits_{G_{a,d}}S^{2}\left(
h,w_{1}\right) \varphi _{\lambda }^{2}dxdt\geq C_{1}\lambda \exp \left(
-3\lambda d^{-\nu }\right) \dint\limits_{G_{a,d}}\left[ \left( \nabla
h\right) ^{2}+h^{2}\right] \varphi _{\lambda }^{2}dxdt
\end{equation*}%
\begin{equation*}
-C_{1}\exp \left( -3\lambda d^{-\nu }\right) \dint\limits_{G_{a,d}}\left(
\nabla h\right) ^{2}\varphi _{\lambda }^{2}dxdt-C_{1}\exp \left( -\lambda
d^{-\nu }\right) \dint\limits_{\xi _{a,d}}\left[ h_{t}^{2}+\left( \nabla
h\right) ^{2}+h^{2}\right] d\sigma
\end{equation*}%
\begin{equation*}
\geq C_{1}\lambda \exp \left( -3\lambda d^{-\nu }\right)
\dint\limits_{G_{a,d}}\left[ \left( \nabla h\right) ^{2}+h^{2}\right]
\varphi _{\lambda }^{2}dxdt-C_{1}\exp \left( -\lambda d^{-\nu }\right)
\dint\limits_{\xi _{a,d}}\left[ h_{t}^{2}+\left( \nabla h\right) ^{2}+h^{2}%
\right] d\sigma
\end{equation*}%
\begin{equation*}
\geq C_{1}\exp \left( 2\lambda q\right) \dint\limits_{G_{a,d-\varepsilon }} 
\left[ \left( \nabla h\right) ^{2}+h^{2}\right] dxdt-C_{1}\exp \left(
-\lambda d^{-\nu }\right) \dint\limits_{\xi _{a,d}}\left[ h_{t}^{2}+\left(
\nabla h\right) ^{2}+h^{2}\right] d\sigma .
\end{equation*}%
Thus, we have established that%
\begin{equation}
\exp \left( -3\lambda d^{-\nu }\right) \dint\limits_{G_{a,d}}S^{2}\left(
h,w_{1}\right) \varphi _{\lambda }^{2}dxdt  \label{4.6}
\end{equation}%
\begin{equation*}
\geq C_{1}\exp \left( 2\lambda q\right) \dint\limits_{G_{a,d-\varepsilon }} 
\left[ \left( \nabla h\right) ^{2}+h^{2}\right] dxdt-C_{1}\exp \left(
-\lambda d^{-\nu }\right) \dint\limits_{\xi _{a,d}}\left[ h_{t}^{2}+\left(
\nabla h\right) ^{2}+h^{2}\right] d\sigma .
\end{equation*}%
Next, since $\alpha \in \left( \exp \left( -\lambda /\left( 2d^{\nu }\right)
\right) ,1\right) ,$ then 
\begin{equation}
-C_{1}\exp \left( -\lambda d^{-\nu }\right) \dint\limits_{\xi _{a,d}}\left[
h_{t}^{2}+\left( \nabla h\right) ^{2}+h^{2}\right] d\sigma \geq -\frac{%
\alpha }{2}\left\Vert h\right\Vert _{H^{4}\left( G_{a,d}\right) }^{2}.
\label{4.7}
\end{equation}%
Combining (\ref{4.3}) with (\ref{4.6}) and (\ref{4.7}), we obtain the target
estimate (\ref{3.19}). $\square $

\end{document}